\pdfoutput=1
\documentclass[conference]{IEEEtran}

\usepackage{amsmath,amssymb}
\usepackage{graphicx}
\usepackage{booktabs}
\usepackage{stfloats}
\usepackage{cite}
\usepackage{xurl}
\usepackage[hidelinks,breaklinks]{hyperref}
\graphicspath{{fig/}}
\usepackage{eso-pic}
\AddToShipoutPictureBG*{\AtPageLowerLeft{\hspace*{0.625in}\raisebox{0.32in}{\parbox{7.25in}{\centering\scriptsize
Accepted for publication in the 2027 IEEE Symposium Series on Computational Intelligence (SSCI 2027).\\
\copyright~2027 IEEE. Personal use of this material is permitted. Permission from IEEE must be obtained for all other uses, in any current or future media, including reprinting/republishing this material for advertising or promotional purposes, creating new collective works, for resale or redistribution to servers or lists, or reuse of any copyrighted component of this work in other works.}}}}

\begin{document}

\title{Field Life: Particle Life as Interacting Chemical Fields}

\author{\IEEEauthorblockN{Iliya Zhechev}
\IEEEauthorblockA{\textit{Sofia University}\\
Sofia, Bulgaria\\
iliya.zhechev@gmail.com}}
\maketitle

\begin{abstract}
Field Life is a cellular automaton inspired by the interaction dynamics of Particle Life. In Particle Life, each particle has a colour, and an interaction matrix sets how strongly each colour attracts or repels each other colour. In Field Life, each colour is a channel of the automaton, and each cell holds a concentration of every colour. One step of the rule has three parts. Each colour is convolved with a ring-shaped kernel, which measures how much of that colour surrounds a cell. The interaction matrix combines these measurements into an attraction for each colour. The mass of each colour then moves toward higher attraction, in a way that conserves mass. A random matrix and a random initial state are usually enough to produce moving structure: bodies that glide, bodies that shed diffuse clouds, and gliders that break up into smaller gliders. We sample rules at random, keep the rules that move and form structure, and describe the patterns that appear: colours that sort into regions, colours that form stripes, and gliders, compact bodies that travel across the grid. The simulation and 120 saved rules, which we call worlds, are released.
\end{abstract}


\section{Introduction}

\begin{figure*}[!t]
  \centering
  \includegraphics[width=0.85\textwidth]{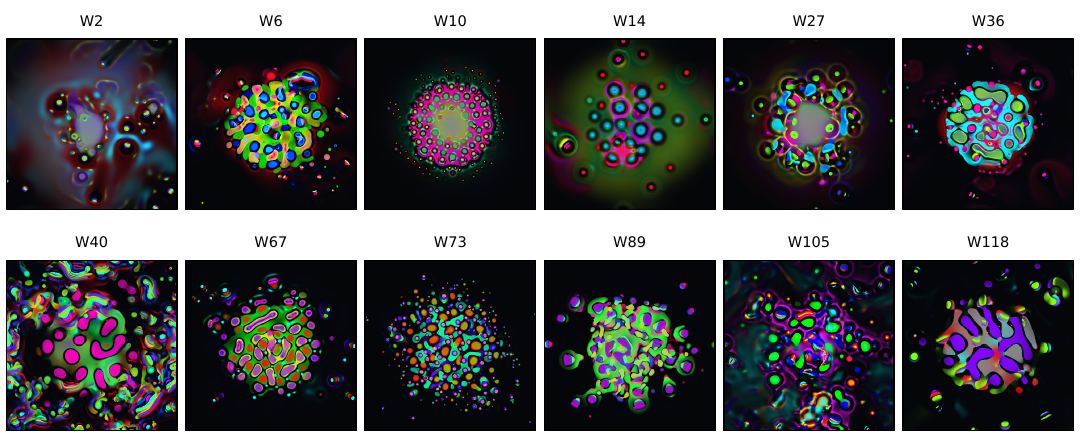}
  \caption{Twelve of the 120 released worlds after 2000 steps, each started from a disc of mixed colours in the centre of an empty $512\times512$ grid (the disc initial state of Section~\ref{sec:explore}). Brightness shows the total concentration and the hue mixes the colours present. Each panel is labelled with the number of its world.}
  \label{fig:gallery}
\end{figure*}

Particle Life \cite{ventrella2017clusters,mohr2022particlelife} is a particle simulation in which particles of different colours attract and repel one another. Two particles that come closer than a short distance push each other away. Beyond that distance, the force between them depends on their colours: an interaction matrix holds one signed strength for every ordered pair of colours, so red may be attracted to green while green is repelled by red. A random matrix is usually enough to produce clusters that chase one another and glide across the plane.

That behaviour has a computational cost. At every step, each particle has to find the particles within its range. The simplest way is to test every pair, at a cost that grows with the square of the number of particles; the usual way is a spatial data structure, a grid of bins or neighbour lists \cite{verlet1967}, which must be maintained as the particles move. A cellular automaton has no such search. It is a dynamical system on a lattice, usually a square grid. Each lattice site is a cell with a state. At each discrete time step, a local rule updates every cell from the cell itself and its neighbours. This value can be binary, as in the Game of Life, where a cell is alive or dead, or a vector of real numbers, as in multi-channel Lenia and in Field Life, where a cell holds one concentration per colour. The neighbours of a cell are fixed by the lattice. Field Life is such an automaton: every cell is updated by the same rule, which consists of convolutions with fixed kernels and a transfer of mass between neighbouring cells.

In Field Life, each colour of Particle Life becomes a concentration field on the grid, and the interaction matrix is kept as it is. The matrix defines how the colours push and pull one another, and the concentrations move under these forces, without mass being created or destroyed. Because the colours spread and drift like concentrations of chemicals, we call them chemical fields.

Random rules in this model produce varied behaviour. Fig.~\ref{fig:gallery} shows twelve worlds, each started from a disc of mixed colours: in some, the colours sort into regions; in some, one colour forms stripes; and in many, compact bodies form and travel. Section~\ref{sec:patterns} follows such bodies; they are gliders in the sense of the Game of Life, patterns that move across the grid as a whole. Bodies of this kind can be lifted out of their world and run alone; some keep moving for thousands of steps, and some break up into smaller gliders.

The question we study is how non-reciprocal interactions and mass conservation together shape the moving structures of such a model. The paper makes four contributions. It defines Field Life, a mass-conserving automaton in which an interaction matrix acts on ring-kernel potentials (Section~\ref{sec:model}). It explores random rules of the model, with the measurements used to keep them (Section~\ref{sec:explore}). It describes the patterns that appear, and shows that the gliders need an asymmetric matrix (Section~\ref{sec:patterns}). It releases the simulation, a reference implementation and 120 worlds. Section~\ref{sec:related} reviews related work, and Section~\ref{sec:discussion} discusses what the model shows and where it falls short.

The simulation runs in the browser at \url{https://www.izhv.eu/field-life/paper-demo/}, with the worlds and gliders numbered as in this paper; the dynamics are far clearer in motion than in still frames. The source, a reference implementation and the 120 worlds are available at \url{https://github.com/ichko/field-life}.

\section{Related work}
\label{sec:related}

Clusters \cite{ventrella2017clusters} and Particle Life \cite{mohr2022particlelife} are simulations in which the force between two coloured particles depends on their colours through a matrix, with a repulsion at short range that applies to every pair. They belong to the line of flocking models that runs from boids \cite{reynolds1987boids} through the Vicsek model \cite{vicsek1995} to Swarm Chemistry \cite{sayama2009swarm}, in which each species has its own motion parameters. In Clusters and Particle Life the clusters move because the matrix is asymmetric: one colour is drawn to another that is repelled by it. Neither model has a peer-reviewed description, so we cite the authors' pages.

SmoothLife \cite{rafler2011smoothlife} moves the Game of Life \cite{gardner1970life} onto a continuous field, and Lenia \cite{chan2019lenia} generalises it with smooth kernels and growth functions. Later work adds several kernels and channels \cite{chan2020expanded}, learned rules \cite{mordvintsev2020nca}, and creatures with sensorimotor behaviour \cite{hamon2022sensorimotor}. In all of these a growth term adds mass to a cell or removes it. Flow-Lenia \cite{plantec2023flowlenia} moves mass along the gradient of the growth term instead, so that mass is conserved. MaCE \cite{papadopoulos2025mace} is a general mass-conserving update: each cell shares its content among its neighbours in proportion to the exponential of an affinity. Field Life keeps this update and replaces the affinity. Multi-channel Lenia and Flow-Lenia also weight ring-kernel convolutions of every channel with signed weights, but pass the sum through a growth function. Field Life has no growth function, moves mass with the MaCE update, and adds a crowding term; what is new is this construction as a whole and the behaviour that follows from it. Without a growth function, the matrix acts on these convolutions directly, so each entry keeps the meaning it has in Particle Life: how strongly one colour is drawn to another. This is what lets us test the role of asymmetry by making the matrix symmetric (Section~\ref{sec:patterns}). Particle Lenia \cite{mordvintsev2022particlelenia} is a particle-based variant of Lenia.

In aggregation models, individuals attract one another at a distance and repel up close, and form clumps and rings \cite{topaz2006nonlocal,kolokolnikov2011ring}. An interaction is non-reciprocal when one species, or type of particle, responds to another differently from how the other responds to it. Mixtures with non-reciprocal interactions form patterns that travel; mixtures with reciprocal ones come to rest \cite{ivlev2015nonreciprocal,agudo2019active,saha2020nrch,fruchart2021,you2020nonreciprocity,brauns2024conserved}. In Field Life each colour is a species, and an asymmetric matrix makes the interactions non-reciprocal.

\section{The model}
\label{sec:model}

\begin{figure*}[!t]
  \centering
  \includegraphics[width=0.9\textwidth]{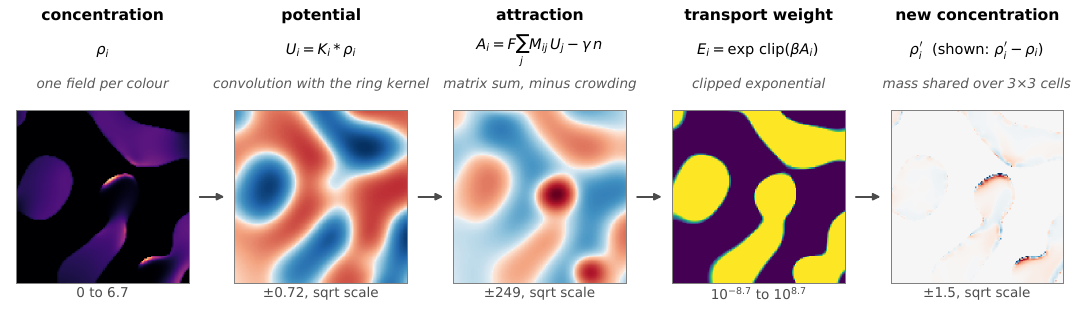}
  \caption{One step of the rule, shown for one colour of world W118 (Fig.~\ref{fig:gallery}) after 2000 steps, in a window 128 cells wide. Each panel is labelled with the quantity it shows and the operation that produces it from the previous one. The potential $U_i$ is negative inside the regions of the colour and positive in a band around them. The attraction $A_i$ is the weighted sum of the potentials of all four colours minus the crowding term. The transport weight $E_i$ runs from $e^{-20}$ to $e^{20}$, seventeen orders of magnitude, and is flat inside the regions, where the exponent is limited. The change $\rho_i' - \rho_i$ sums to zero. Signed panels use a square-root colour scale; each panel states its value range.}
  \label{fig:model}
\end{figure*}

\textbf{State.} The grid is square, $L \times L$ cells, with periodic boundaries. There are $C$ colours. Each colour $i$, from 1 to $C$, is a field $\rho_i$: one non-negative number per cell, the concentration of colour $i$ in that cell. The mass of colour $i$ is the sum of $\rho_i$ over the grid. One step of the rule computes, for each colour, a potential, then an attraction, then a move (Fig.~\ref{fig:model}).

\begin{figure}[t]
  \centering
  \includegraphics[width=0.95\columnwidth]{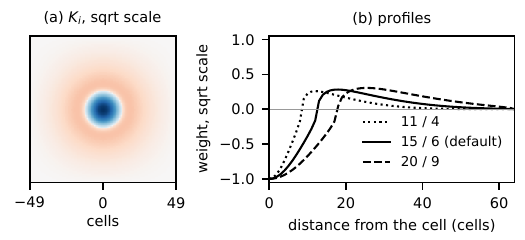}
  \caption{The ring kernel. (a) The disc of weights for the default widths, $\sigma_w = 15$ and $\sigma_n = 6$ cells; red is positive, blue negative. (b) Weight against distance from the cell for a narrow, the default and a wide ring, each divided by the magnitude of its central value; the labels give $\sigma_w$ / $\sigma_n$ in cells. Both panels use a square-root scale, so that the low outer ring is visible.}
  \label{fig:kernels}
\end{figure}

\textbf{Potential.} Each colour $i$ has a kernel $K_i$, a disc of weights centred on the cell being updated. The kernel is ring-shaped: negative at the centre and positive on a ring around it (Fig.~\ref{fig:kernels}). The potential of colour $i$ is the convolution of its field with its kernel,
\begin{equation}
  U_i = K_i * \rho_i .
  \label{eq:potential}
\end{equation}
The potential is high at a cell that colour $i$ surrounds at the distance of the ring, and low at a cell that colour $i$ occupies.

\begin{figure*}[!t]
  \centering
  \includegraphics[width=0.8\textwidth]{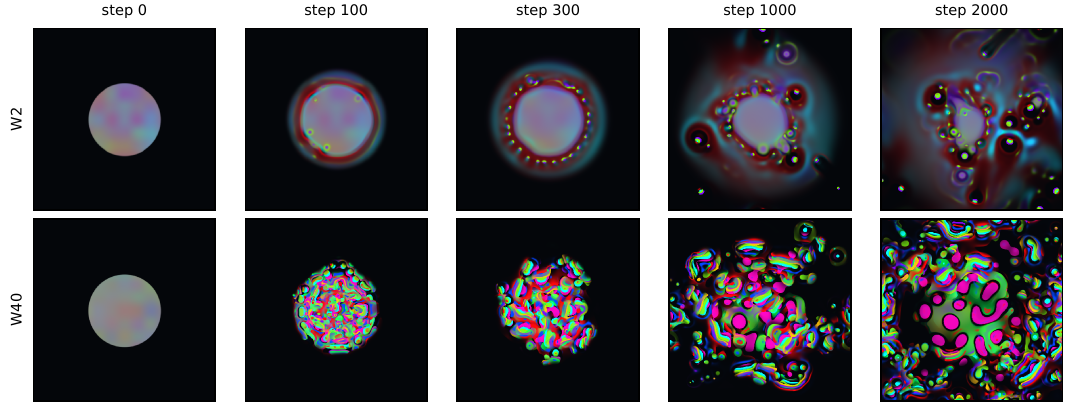}
  \caption{Two worlds developing from a disc of mixed colours, at steps 0, 100, 300, 1000 and 2000. Top: W2, where a mixed core remains. Bottom: W40, where the disc is gone by step 1000. In both, the edge of the disc sorts into one-colour patches first, patches move off, and the sorting works inward; drawn as in Fig.~\ref{fig:gallery}.}
  \label{fig:rollouts}
\end{figure*}

\textbf{Attraction.} The interaction matrix $M$ holds one signed weight $M_{ij}$ for each ordered pair of colours: how colour $i$ responds to colour $j$. The attraction of colour $i$ at a cell is
\begin{equation}
  A_i = F \sum_{j=1}^{C} M_{ij}\, U_j \;-\; \gamma\, n, \qquad n = \sum_{j=1}^{C} G_j * \rho_j ,
  \label{eq:attraction}
\end{equation}
where $F$ is the force, one number that scales the whole matrix term, and $n$ is the crowding: the total mass of all colours near the cell, obtained by blurring each field $\rho_j$ with a narrow Gaussian $G_j$ and summing. The crowding enters with weight $\gamma$, the repulsion. A positive $M_{ij}$ draws colour $i$ to cells at ring distance from colour $j$ and away from cells on top of it; a negative $M_{ij}$ does the opposite. When we say below that colour $i$ is attracted to or repelled by colour $j$, we mean this: attraction to the ring around $j$, or away from it. The crowding pushes mass out of cells that are already full, whatever their colours; it is the short-range repulsion of Particle Life, which acts between every pair of particles.

\textbf{Motion.} Mass moves toward higher attraction by the MaCE update \cite{papadopoulos2025mace}. Each cell has a transport weight
\begin{equation}
  E_i = \exp\!\big(\mathrm{clip}(\beta A_i, -20, 20)\big),
  \label{eq:weight}
\end{equation}
where the limit on the exponent is explained under Implementation, and $\beta$ is the sharpness, MaCE's own parameter: larger $\beta$ sends more of a cell's mass to the neighbour with the highest attraction. Every cell distributes its mass among the nine cells of its $3\times3$ neighbourhood $\mathcal{N}$, itself included, in proportion to their weights:
\begin{equation}
  \rho_i'(x) = \sum_{y \in \mathcal{N}(x)} \rho_i(y)\, \frac{E_i(x)}{\sum_{z \in \mathcal{N}(y)} E_i(z)} .
  \label{eq:transport}
\end{equation}
The fraction is the share of the mass of cell $y$ that goes to cell $x$; its denominator runs over the neighbourhood of the sender $y$. The nine shares of every cell sum to one, so every unit of mass that leaves a cell arrives in a neighbouring cell, and the mass of each colour is conserved, up to rounding. A world is specified by the matrix $M$, one kernel per colour, and the three scalars $F$, $\gamma$ and $\beta$.

\textbf{Implementation.} The kernel of colour $i$ is a wide Gaussian of standard deviation $\sigma_w$ minus a narrow Gaussian of standard deviation $\sigma_n$. Each Gaussian is cut at 3.2 times the largest $\sigma_w$ of the world and normalised to unit mass, so the weights of the difference sum to zero; the difference is scaled so that its absolute values sum to one. $G_j$ is the unit-mass narrow Gaussian of colour $j$ before this division, so $\sigma_n$ of colour $j$ sets both the core of its kernel and its crowding blur. The defaults are $\sigma_w = 15$ and $\sigma_n = 6$ cells, $F = 90$, $\gamma = 6$ and $\beta = 3$; we set them by hand while watching the simulation. Before the exponential in (\ref{eq:weight}), $\beta A_i$ is limited to $\pm 20$, which keeps the exponential within the range of a 32-bit float; we set the value by hand. With the default scalars the exponent is beyond the limit over most of a body, so inside a body the nine weights are equal and the mass spreads evenly; the sign of the attraction acts at the edge of a body, where it decides whether the edge advances or retreats. The world is $512 \times 512$ cells. The simulation runs in a web browser through WebGL2 at the display rate, 120 steps per second on an Apple M4 Max laptop. The two Gaussians are separable, so each convolution is two one-dimensional blurs. A NumPy implementation of (\ref{eq:potential})--(\ref{eq:transport}) reproduces one step of the browser implementation to within a rounding error of $10^{-6}$ per cell, and after 300 steps the two implementations place the centre of mass of a glider within $10^{-3}$ cells of each other. The relative drift of the total mass over 300 steps is below $10^{-5}$ in 32-bit arithmetic and $5\times10^{-14}$ in 64-bit. The images in this paper are rendered from simulation states with the colour mapping of the simulation: brightness from the total concentration, hue from the colours present; colour $i$ has the hue at fraction $(i-1)/C$ of the colour wheel, so with four colours, colours 1 to 4 are red, green, cyan and purple.

\section{Exploring the parameter space}
\label{sec:explore}

\begin{figure*}[!t]
  \centering
  \includegraphics[width=0.82\textwidth]{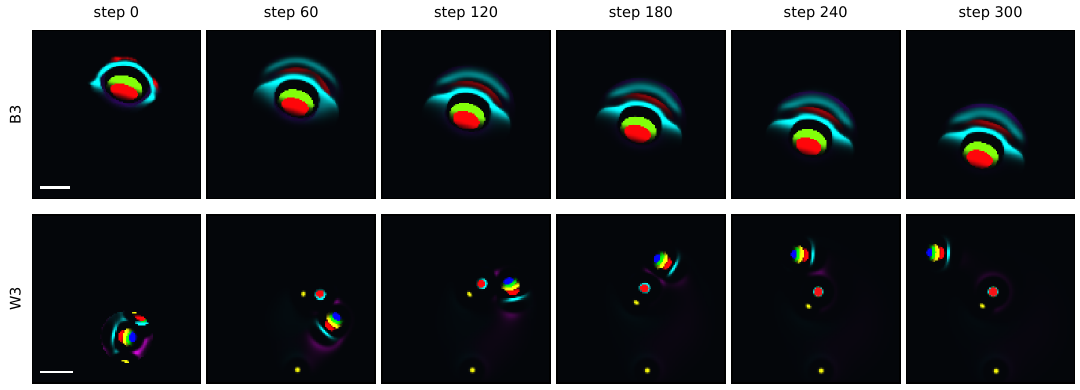}
  \caption{Two bodies, each alone on an empty grid under the rule of its parent world, at steps 0, 60, 120, 180, 240 and 300. The window of each row is fixed, so the body moves across it; the bar in the first panel of each row is 20 cells long. Top: B3, which travels in a straight line and keeps its shape; colours 1 to 4 are drawn red, green, cyan and purple, and the hues mix where the colours overlap. Bottom: a body cut out of world W3 after 1000 steps from a disc, an example of a body that changes as it travels; it is not one of the seven gliders of Table~\ref{tab:creatures}. It leaves a dot of one colour behind. Both parent worlds can be watched in the browser demo.}
  \label{fig:motion}
\end{figure*}

\textbf{Method.} We draw a rule at random, run it from a random initial state for 300 steps, and keep it if it moves and forms structure. The test uses three measurements, each computed on the total concentration field, the sum of all colours. Activity is the fraction of mass that changes from one step to the next, $\sum_x |\rho_t(x) - \rho_{t-1}(x)| / \sum_x \rho_t(x)$, averaged over the run; it stands for a world that is still evolving rather than frozen. Structure size is the typical extent of the 4-connected regions above a tenth of the peak concentration at the last step, the mass-weighted median of $\sqrt{\rm area}$ over those regions; it stands for mass gathered into bodies rather than noise or haze. Local displacement is computed on the non-overlapping $32\times32$ patches of the field. For each patch it is the shift of up to four cells that minimises the sum of absolute differences between the patch and the previous step; the patch displacements are averaged over the patches, weighted by the mass that changed in each patch, and over the run. It stands for bodies that travel as wholes rather than churn in place. These measurements are modelled on Lenia's statistical descriptors \cite{chan2019lenia} but are not the same quantities.

\textbf{Details.} All values in this paragraph, the sampling ranges, the initial states and the thresholds, were chosen by hand: they are the settings under which the simulations looked interesting to us, and none is derived from theory. The model allows any number of colours; the browser implementation supports up to 8, and a sample has $C$ colours, with $C$ drawn from 4 to 8. The entries of $M$ are drawn uniformly from $[-1, 1]$. Each colour receives a ring with $\sigma_w$ drawn uniformly between 9 and 22 cells and $\sigma_w / \sigma_n$ drawn uniformly between 1.7 and 5.7; these are the ranges of the simulation's kernel sampler. The scalars keep their defaults in the counted run below. The 120 released worlds were sampled more widely: a global scale on both widths, so that their $\sigma_w$ ranges from 6 to 33 cells, and in 52 of them the scalars as well ($F$ from 37 to 188, $\gamma$ from 2 to 13, $\beta$ from 1.8 to 5.3). The initial state is either noise or a disc. Noise: each colour starts as random patches spread over the whole grid, with a mass of $0.12\,L^2$; the released code gives the recipe. Disc: a disc of radius $0.2\,L$ in the centre of an empty grid, with every colour present in every cell of the disc at a concentration between 0.3 and 0.7 that varies smoothly from cell to cell (independent noise for each colour, on grids of 6 and 14 points), and an edge that fades to zero over 6 cells. A sample is kept when its activity exceeds 0.02, its structure size exceeds 9 cells (so that scattered specks do not count as structure), and its displacement exceeds 0.05 cells per step. Runs last 300 steps, which is enough for structure to form from noise in every world we tried.

\textbf{Result.} We ran the sampler four times with different random seeds (0 to 3), 12 rules each, 48 in all, from noise. Of these, 44 pass (12, 11, 10 and 11 per run). In 30 of the 44 (9, 8, 4 and 9), a stricter test for compact, single-coloured moving bodies, part of the released sampler, also passes: displacement above 0.25 cells per step; fewer than 40\% of cells occupied, where a cell is occupied when its total concentration is above a tenth of the peak; and purity above 0.6, where the purity of a cell is $\sum_i \rho_i^2 / (\sum_i \rho_i)^2$, equal to 1 when the cell holds one colour, and the purity of a sample is the mass-weighted mean over cells. The pass rate depends on the sampling ranges, which were chosen by hand. We picked the 120 released worlds by eye, while watching the simulation, from the same kind of sampling. Fig.~\ref{fig:gallery} shows twelve of them.

\section{The patterns}
\label{sec:patterns}

\textbf{From a disc.} We start from the disc of Section~\ref{sec:explore}: a round patch in which every cell holds every colour in random amounts, on an empty grid. Without the randomness the disc would be uniform and would stay uniform. When the rule runs, at first only the edge of the disc changes. At the edge, cells that hold all colours come to hold one colour. These one-colour patches begin to move, and some break off and travel into the empty space. The same change works deeper and deeper into the disc. In some worlds the disc is gone after about a thousand steps, and what is left is one-colour regions and moving bodies; in others a mixed core remains for thousands of steps while the sorted rim moves around it. Fig.~\ref{fig:rollouts} shows one of each.

\textbf{Sorting and stripes.} Sorting is what happens to the disc: a cell that holds several colours ends up holding one, and the world divides into one-colour regions. Many of the released worlds sort, from a disc and from noise alike (W6, W14, W40, W105 and W118 in Fig.~\ref{fig:gallery}). Stripes are a second pattern: one colour settles into parallel bands, all the same distance apart, about the radius of the ring of the kernel. The ring kernel gives its highest weight to mass at that distance, and bands at that spacing place mass at that distance from every cell at once. A few of the released worlds do this (W14, W36, W73 and W118 in Fig.~\ref{fig:gallery}). The same bands appear in reaction--diffusion systems \cite{turing1952,pearson1993}.

\begin{table}[t]
  \caption{The seven gliders: parent world, its number of colours $C$ and kernel width $\sigma_w$ (the range over colours where they differ), and the size and speed of the body, defined in the text.}
  \label{tab:creatures}
  \centering
  \scriptsize\setlength{\tabcolsep}{3.5pt}\renewcommand{\arraystretch}{0.9}
  \begin{tabular}{@{}lccccc@{}}
\toprule
Body & World & $C$ & $\sigma_w$ (cells) & Size (cells) & Speed (cells/step) \\
\midrule
B0 & W1 & 4 & 16--21 & 12$\times$8 & 0.89 \\
B1 & W33 & 6 & 13--25 & 12$\times$7 & 0.33 \\
B2 & W36 & 4 & 14 & 105$\times$73 & 0.23 \\
B3 & W49 & 4 & 15 & 25$\times$18 & 0.16 \\
B4 & W39 & 8 & 12--26 & 37$\times$33 & 0.14 \\
B5 & W16 & 4 & 17--22 & 39$\times$29 & 0.06 \\
B6 & W71 & 6 & 21--23 & 15$\times$16 & 0.04 \\
\bottomrule
\end{tabular}

\end{table}

\begin{figure*}[!t]
  \centering
  \includegraphics[width=0.85\textwidth]{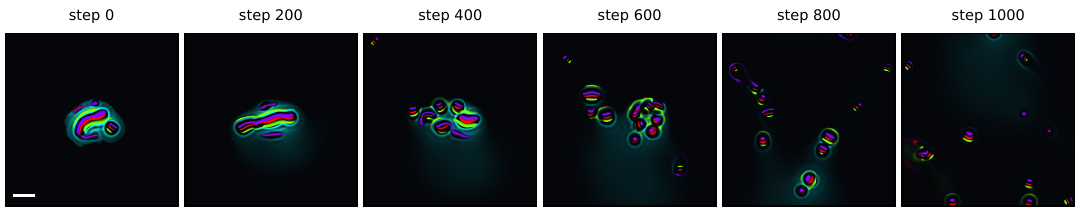}
  \caption{Glider B2 alone on an empty grid for 1000 steps, in a window 360 cells wide that follows the centre of mass. The group breaks into separate bodies, and the bodies into small ring-shaped gliders that travel on their own. The bar is 40 cells long.}
  \label{fig:division}
\end{figure*}

\begin{figure*}[!b]
  \centering
  \includegraphics[width=\textwidth]{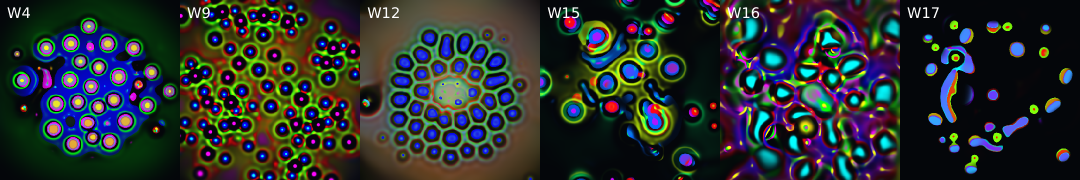}
  \caption{Six more released worlds, to show the range of patterns the rule produces, each after 2000 steps from a disc and drawn as in Fig.~\ref{fig:gallery}; each panel is a $384\times384$ window around the centre of the world. All 120 run in the browser demo.}
  \label{fig:appendix}
\end{figure*}

\textbf{Gliders.} A glider is a body, or a group of bodies with a common heading, that keeps travelling; its shape may change as it goes. To test a body, we copy it onto an empty grid under the rule of its own world and run it for 300 steps; it is a glider if its centre of mass advances by more than 10 cells in those 300 steps. We have seven, B0 to B6, each from a different world; the slowest passes the test with a margin of 2 cells. Table~\ref{tab:creatures} gives their parent world, number of colours, kernel width, size and speed. Size is the width and height, in grid cells, of the box around the part of the body above a tenth of its peak concentration at step 0; speed is the distance travelled by the centre of mass per step over the 300 steps. Fig.~\ref{fig:motion} follows B3 for 300 steps in a fixed window, together with a body cut out of world W3 in the same way. All seven move in nearly straight lines, but not all along the lattice: B1, B5 and B6 head along a lattice axis, while B0, B3 and B4 head at about $110^\circ$, $105^\circ$ and $55^\circ$ from the $x$ axis of the grid. When the starting body of B0 or B3 is rotated by $45^\circ$, its heading turns by $48^\circ$--$50^\circ$, so the direction of motion is set by the body, not by the lattice. Over three runs from starts with $\pm10\%$ noise added to every cell, the speed of each glider varies by at most 0.01 cells per step, and its heading by at most $2^\circ$, except for B2, which breaks up. B3 and B0 keep their speed and their size over 3000 steps. B6, the slowest, sheds mass into a faint cloud around it, 12\% of its total by step 3000, while its core travels on. B0 and B1 are each a crescent of one colour with a dot of a second colour ahead of it; B3 and B4 are each a shell of one colour around a core of a second, with a third trailing behind; B5 and B6 are domes of two stacked colours that travel flat side first; B2 is a loose group of tubes and rings that rearranges as it travels while keeping a common heading; it passes the test as a group. Run for longer, it breaks up: by step 500 it is a handful of bodies, and by step 1000 a dozen small ring-shaped gliders that travel on their own (Fig.~\ref{fig:division}).

\textbf{Asymmetry.} When the matrix of each glider is replaced by its symmetric part $(M + M^\top)/2$, so that every pair of colours pulls or pushes equally, and the body is run alone on an empty grid from its saved state, all seven stop within 150 steps and settle into round, motionless patterns; their speeds fall from 0.04--0.89 to below 0.02 cells per step. The same happens when, in addition, every colour is given the default kernel. This is what happens in Particle Life when the matrix is made symmetric, and it is what the theory of non-reciprocal interactions predicts \cite{ivlev2015nonreciprocal,agudo2019active,saha2020nrch,fruchart2021,you2020nonreciprocity,brauns2024conserved}. With a symmetric matrix and one shared kernel, the attraction is the derivative of an energy, the mass moves so as to lower it, and the system relaxes; a pair in which one colour is drawn to another that is repelled by it has no such energy and can move indefinitely. The argument is exact only for a shared kernel, which five of the seven gliders do not have; they stop all the same. As a control, we kept the symmetric part of each matrix and replaced the antisymmetric part by a random one with the same mean absolute entry, twice per glider: in 8 of the 14 runs the body moves faster than 0.02 cells per step, in one (B5) faster than under its own matrix, and in the other 6 it stops. Asymmetry is needed for motion, but a given body does not move under just any asymmetric matrix. The matrix of B3 is
\begin{equation*}
  M = \begin{pmatrix}
    -0.27 & \phantom{-}0.16 & -0.90 & -0.88 \\
    -0.61 & \phantom{-}0.09 & -0.29 & \phantom{-}0.65 \\
    \phantom{-}0.84 & \phantom{-}0.00 & \phantom{-}0.65 & \phantom{-}0.16 \\
    \phantom{-}0.49 & \phantom{-}0.70 & -0.04 & -0.16
  \end{pmatrix},
\end{equation*}
with colours 1 to 4 drawn red, green, cyan and purple in Fig.~\ref{fig:motion}. Cyan (colour 3) is attracted to red (colour 1), $M_{31} = 0.84$, and red is repelled by cyan, $M_{13} = -0.90$: the cyan shell settles at ring distance behind the red core, the core moves off that ring, and the shell follows. Purple and green attract each other, $M_{42} = 0.70$ and $M_{24} = 0.65$, stay together, and travel with the body.

\section{Discussion}
\label{sec:discussion}

Three results may be useful to others who build models of this kind. First, motion needs no growth term: mass conservation and an asymmetric matrix give gliders from random rules, and the symmetric part of the same matrix stops them. Second, the limit on the exponent of the MaCE weights is a parameter of the dynamics, not only a numerical safeguard. Run alone with limits of 5, 10, 20, 40 and 80, B3 keeps its speed of 0.16 cells per step up to a limit of 20, while B0 slows from 0.89 to 0.67 cells per step as the limit falls to 5. Above 20 the peak concentration grows, to 2.2--2.5 times its starting value after 300 steps at a limit of 80; B3 then speeds up and B0 breaks up. Third, a $3\times3$ transport does not by itself tie motion to the lattice: rotating a glider rotates its heading.

Each colour has its own kernel, which sets the potential the other colours see of it, and its own row of the matrix, which sets its attraction to each of them; a colour therefore behaves as a species. A glider is an assembly of species. In B3, one pair of colours drives the body, one drawn to the other that is repelled by it, and another pair, attracted to each other, is carried with it.

We picked the 120 released worlds by eye, and the sample of 48 rules on which we report counts is small and drawn from ranges chosen by hand. The glider measurements vary little over perturbed starts, but each glider is one body from one world. The transport is restricted to the $3\times3$ neighbourhood, so no body moves faster than one cell per step, and the model has no continuum limit of the kind Lenia has. We have not compared the dynamics with those of Particle Life under the same matrix, nor the cost of a step with that of a particle implementation.

\section{Conclusion}

Field Life puts the interaction matrix of Particle Life on a grid of concentration fields, with a ring kernel in place of the pairwise force and a mass-conserving move in place of particle motion. Random matrices produce sorted regions, stripes and gliders; the gliders travel in straight lines and can be lifted out of their worlds and run alone on an empty grid. The simulation and the 120 worlds are released; Fig.~\ref{fig:appendix} shows six more of them.

\section*{Acknowledgment}
This paper was prepared with the assistance of AI tools: Anthropic's Claude (several model versions) \cite{anthropic2026claude} and the Claude Code coding agent \cite{anthropic2026claudecode}, used for drafting, language refinement, formatting and code assistance. The author reviewed all AI-assisted text and code, and the experimental results, analysis, and conclusions are the author's own work.

\clearpage
\raggedbottom
\bibliographystyle{IEEEtran}
\bibliography{refs}

\end{document}